\documentclass[aps,prd,reprint, superscriptaddress, nofootinbib]{revtex4-2}

\usepackage{graphicx}        
\usepackage[caption=false, position=top]{subfig}
\usepackage{aas_macros} 
\usepackage{amssymb,amsmath,placeins,epsfig,array,bm}
\usepackage[usenames,dvipsnames]{color}
\usepackage[normalem]{ulem}
\usepackage[breaklinks,colorlinks,urlcolor=blue,citecolor=blue,linkcolor=blue]{hyperref}
\usepackage{float}
\usepackage[T1]{fontenc}
\usepackage{times}
\usepackage{soul}
\usepackage{array,multirow,graphicx}
\usepackage{xcolor}

\begin{document}

\title{A representation learning approach\\to probe for dynamical dark energy in matter power spectra}

\author{Davide Piras}
\email[]{davide.piras@unige.ch}
\affiliation{Département de Physique Théorique, Université de Genève, 24 quai Ernest Ansermet, 1211 Genève 4, Switzerland}
\affiliation{Centre Universitaire d’Informatique, Université de Genève, 7 route de Drize, 1227 Genève, Switzerland}

\author{Lucas Lombriser}
\affiliation{Département de Physique Théorique, Université de Genève, 24 quai Ernest Ansermet, 1211 Genève 4, Switzerland}

\date{\today}

\begin{abstract}
We present \texttt{DE-VAE}, a variational autoencoder (VAE) architecture to search for a compressed representation of dynamical dark energy (DE) models in observational studies of the cosmic large-scale structure. \texttt{DE-VAE} is trained on matter power spectra boosts generated at wavenumbers $k \in (0.01 - 2.5) \ h/$Mpc and at four redshift values $z \in (0.1, 0.48, 0.78, 1.5)$ for the most typical dynamical DE parametrization with two extra parameters describing an evolving DE equation of state. The boosts are compressed to a lower-dimensional representation, which is concatenated with standard cold dark matter (CDM) parameters and then mapped back to reconstructed boosts; both the compression and the reconstruction components are parametrized as neural networks. Remarkably, we find that a single latent parameter is sufficient to predict 95\% (99\%) of DE power spectra generated over a broad range of cosmological parameters within $1\sigma$ ($2\sigma$) of a Gaussian error which includes cosmic variance, shot noise and systematic effects for a Stage IV-like survey. This single parameter shows a high mutual information with the two DE parameters, and these three variables can be linked together with an explicit equation through symbolic regression. Considering a model with two latent variables only marginally improves the accuracy of the predictions, and adding a third latent variable has no significant impact on the model's performance. We discuss how the \texttt{DE-VAE} architecture can be extended from a proof of concept to a general framework to be employed in the search for a common lower-dimensional parametrization of a wide range of beyond\nobreakdash-$\Lambda$CDM models and for different cosmological datasets. Such a framework could then both inform the development of cosmological surveys by targeting optimal probes, and provide theoretical insight into the common phenomenological aspects of beyond\nobreakdash-$\Lambda$CDM models.
\end{abstract}


\maketitle


\section{Introduction}
Despite its remarkable success in correctly reproducing cosmological observations, the concordance $\Lambda$CDM model still carries some unresolved questions. In particular, the nature of its two core components, namely cold dark matter (CDM) and the cosmological constant $\Lambda$, remains a mystery~\citep{Bull16}. For example, assuming $\Lambda$ as a vacuum energy density, it is currently unclear how to reconcile its value as measured by cosmological probes~\citep{Planck20} with estimates from quantum field theory~\citep{Martin12}, invoking the concept of dark energy (DE) as the driver of the late-time accelerated expansion of the Universe. Moreover, in recent years significant tensions have emerged between different measurements of $\Lambda$CDM parameters, which could hint at modifications of the concordance model or could be attributed to experimental systematic effects. These tensions include a ${\sim} 5\sigma$ difference between the constraints on the expansion rate $H_0$ coming from local distance ladder and CMB experiments~\citep{Riess22}, and a potential discrepancy between the values of the structure growth parameter $S_8$ measured with large-scale structure probes and cosmic microwave background (CMB) experiments~\citep{Heymans21, DES23}. Current and upcoming surveys such as the Vera Rubin Observatory\footnote{\href{https://www.lsst.org/}{https://www.lsst.org/}}~\citep{Ivezic19}, \textit{Euclid}\footnote{\href{https://www.euclid-ec.org/}{https://www.euclid-ec.org/}}~\citep{Laureijs11}, the Nancy Grace Roman Space Telescope\footnote{\href{https://roman.gsfc.nasa.gov/}{https://roman.gsfc.nasa.gov/}}~\citep{Spergel15} and the Simons Observatory\footnote{\href{https://simonsobservatory.org/}{https://simonsobservatory.org/}}~\citep{Ade19} promise to provide unprecedented insight on the nature of dark matter and dark energy, and will shed further light on these tensions.

These theoretical issues and observational discrepancies have motivated a broad variety of alternative models to $\Lambda$CDM (we refer the reader to Refs.~\citep{Copeland06, Clifton12, Koyama16, Joyce15, Joyce16, Bull16, Ishak19, DiValentino21} for reviews). The rich manifold of proposals ranges from dynamical dark energy and interacting dark sector models to modifications of general relativity (GR), extra dimensions, and violations of Lorentz invariance, to name just a few. Attempts to address the $\Lambda$CDM tensions with modifications of GR also need to comply with stringent tests on small scales~\citep{Williams04, Schlamminger08, Wagner12, Will14, Peebles05, Ishak19}. On the other hand, such models can exhibit complex effects that act as screening mechanisms and suppress local deviations from GR~\citep{Vainshtein72, Babichev13, Babichev09, Khoury04, Hinterbichler10}; these are however hard to parametrize, due to their nonlinear nature (see Ref.~\cite{Lombriser18} for a review). Traditionally, the impact of the manifold proposals on the matter distribution in the Universe, as measured by the galaxy correlation function or its Fourier counterpart (the power spectrum), has been of particular interest, particularly in light of the aforementioned upcoming large-scale structure surveys~\citep{Hu07b, Zhang07, Amendola08, Bertschinger08, Bean10, Pogosian10, Yong10}.

Typically, extensions to $\Lambda$CDM introduce additional parameters and functions to describe the effect of the proposed modification on cosmological observables. A variety of approaches have been introduced to develop a common parametrization for the vast space of models (see Ref.~\citep{Lombriser18} for a review). In particular, in modified gravity theories the common approach is to assume a phenomenological parametrization framework where the late-time growth is modified with scale- and time-dependent functions, usually dubbed $\mu$ and $\Sigma$, which affect the Poisson equation for the gravitational and the lensing potential, respectively~\citep{Bertschinger08}. In order to agnostically test the plethora of models mapping onto this framework, one must then deal with large parameter spaces, and has to choose a parametric form for the additional functions; in turn, this can lead to suboptimal modeling choices, while any statistical evidence of such a hugely-extended parameter space is washed away by the large extra-dimensionality. For instance, ignoring scale-dependencies such as from Yukawa forces or nonlinear screening effects, only being agnostic about the time dependence of $\mu$ and $\Sigma$ essentially already introduces an infinite number of parameters to constrain. This grows even further if considering dark sector interaction models, where each matter species comes with its own set of $\mu$ and $\Sigma$ functions. Performing Bayesian inference in such high-dimensional scenarios gives rise to parameter degeneracies, as well as being extremely computationally expensive due to the curse of dimensionality, despite recent advances in more efficient sampling techniques like Hamiltonian Monte Carlo \citep{Duane87, Hoffman14}. Even without taking into account systematic and observational effects that must be modeled in large-scale structure analyses, the number of theoretical parameters in $\Lambda$CDM extensions can grow infinitely due to the additional modifications being considered. Therefore, being parsimonious and finding a minimal set of parameters to successfully describe beyond\nobreakdash-$\Lambda$CDM models is imperative.

Along these lines of research, Ref.~\cite{Huterer03} explored a data-driven parameterization of DE models based on principal component analysis (PCA). Subsequent work has developed a similar approach within a Bayesian framework~\citep{Crittenden09}, and applied it in the context of weak lensing, spectroscopic and CMB surveys~\citep{Zhao09, Hojjati12, Shinsuke13}. More recently, machine learning algorithms have been proposed to distinguish between different cosmological models. Ref.~\cite{Schmelzle17} developed a deep convolutional neural network (CNN) that can discriminate between five $\Lambda$CDM models with different values of matter abundance $\Omega_{\textrm{m}}$ and matter fluctuations $\sigma_8$ given noisy convergence mass maps, with a better performance than skewness and kurtosis. Ref.~\cite{Peel19} trained a CNN on convergence maps to classify between different $f(R)$ gravities with massive neutrinos and $\Lambda$CDM, while Ref.~\cite{Mancarella22} developed the Bayesian Cosmological Network (\texttt{BaCoN}), a Bayesian classifier that was trained to discriminate between either five different cosmological models, or between $\Lambda$CDM and non-$\Lambda$CDM, given their predicted matter power spectrum at different redshifts with 95\% accuracy. More generally, machine learning algorithms have shown great capabilities at extracting and compressing information from large or high-dimensional cosmological datasets~\citep{Portillo20, Sedaghat21, Sarmiento21, Teimoorinia22, LucieSmith22, LucieSmith23}

In this paper, we develop a variational autoencoder (VAE) to find a low-dimensional representation of a dynamical dark energy model with two additional free parameters with respect to $\Lambda$CDM; we focus on a single parametrization model of dynamical DE as a proof of concept, with the aim of extending our approach to a larger set of beyond\nobreakdash-$\Lambda$CDM models in future work. We train the network on the theoretical matter power spectrum boost\footnote{While the power spectrum is not directly observable, it serves as an observational proxy for our proof-of-concept study, given its aforementioned importance in cosmological analyses.} with respect to the $\Lambda$CDM prediction, and design its architecture and loss function to encourage a disentangled representation of the extra parameters. We investigate how many latent variables are needed to reconstruct the power spectra with sufficient accuracy, and investigate the links between the latents and the known additional parameters through the information-theoretic metric of mutual information (MI) and symbolic regression (SR). The latter returns a human-readable equation linking cosmological parameters and latent variables, which can then inform the development of theoretical works and new cosmological observables.

The paper is structured as follows. In Sect.~\ref{sec:data} we describe the power spectrum boosts that we use to train the machine learning model, which we present in detail in Sect.~\ref{sec:model}. In Sect.~\ref{sec:results}, we show the performance of the model with varying number of latent variables, and use both MI and SR to extract insight from the latent space. We discuss our results in Sect.~\ref{sec:discuss}, and draw our conclusions in Sect.~\ref{sec:conclusions}.

\begin{figure*}
	\centering
	\includegraphics[width=\textwidth, trim={0.0cm 0.cm 0.cm 0cm},clip]{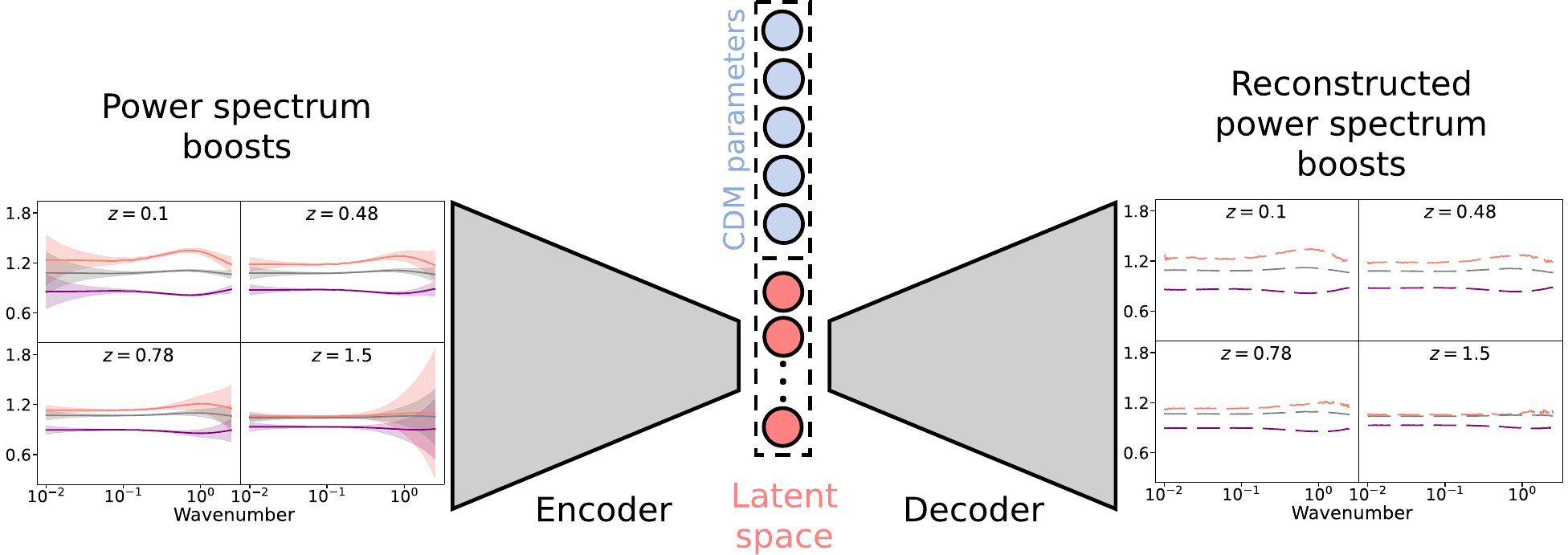}
	\caption{Schematic view of the \texttt{DE-VAE} model employed in this work. The input data consist of power spectrum boosts, as defined in Eq.~(\ref{eq:boost}), produced at four different redshift bins and for wavenumbers $k \in (0.01 - 2.5) \ h/$Mpc. These boosts are compressed by an encoder, parametrized with convolutional neural networks, to a set of disentangled latent variables (in red). These variables are then concatenated with CDM parameters (in blue) and transformed back to power spectrum boosts through the decoder. All model details are reported in Sect.~\ref{sec:model}.}
	\label{fig:scheme}
\end{figure*} 

\section{Data}
\label{sec:data}
As highlighted in the introduction, deviations from the $\Lambda$CDM model modify the expected matter distribution in the Universe. Consequently, a particular quantity of interest given forthcoming surveys is the matter power spectrum (the Fourier transform of the 2-point matter correlation function), computed at different wavenumbers $k$ and redshifts $z$. We focus on the power spectrum boost $\mathcal{B}(k, z)$, defined as
\begin{equation}
    \mathcal{B}(k, z) = \frac{P^{\textrm{b\nobreakdash-}\Lambda\textrm{CDM}}_{\delta\delta} (k, z)}{P^{\Lambda\textrm{CDM}}_{\delta\delta} (k, z)} \ ,
    \label{eq:boost}
\end{equation}
where $P^{\Lambda\textrm{CDM}}_{\delta\delta} (k, z)$ is the matter power spectrum predicted by $\Lambda$CDM (with constant dark energy density), while $P^{\textrm{b\nobreakdash-}\Lambda\textrm{CDM}}_{\delta\delta} (k, z)$ represents the corresponding quantity in a beyond\nobreakdash-$\Lambda$CDM scenario. As a proof of concept, in this work we focus on dynamical dark energy with time-varying equation of state, adopting the most common parametrization~\citep{Chevallier01, Linder03}
\begin{equation}
    w(a) = w_0 + (1-a)w_a \ ,
\end{equation}
where $a$ is the scale factor, which introduces two extra parameters: $w_0$ and $w_a$. We refer to this model as $w_0w_a$CDM; $\Lambda$CDM is recovered for $w_0 = -1$ and $w_a=0$, with current constraints being statistically consistent with the $\Lambda$CDM values of these parameters~\cite{Menci20, Abbott23}. We reiterate that, while in this work we only focus on a single beyond\nobreakdash-$\Lambda$CDM model, in future work we plan to train our proposed framework with multiple classes of non-standard models, as well as with different cosmological probes.

We generate all matter power spectra using the Code for Anisotropies in the Microwave Background (CAMB,~\cite{Lewis11}), using \texttt{HMCode} to compute the nonlinear corrections~\cite{Mead15, Mead16, Mead21}, in 400 bins in the $k$-range $(0.01 - 2.5) \ h/$Mpc and for four redshift values $z \in (0.1, 0.48, 0.78, 1.5)$, motivated by a \textit{Euclid}-like survey. We vary five CDM cosmological parameters: the matter ($\Omega_{\textrm{m}}$) and baryon energy density ($\Omega_{\textrm{b}}$), the dimensionless Hubble parameter $(h)$, the primordial amplitude $(A_{\textrm{s}})$ and the scalar spectral index $(n_{\textrm{s}})$. We consider a uniform range for all cosmological parameters, with the intervals reported in Table~\ref{tab:priors}. These correspond to five standard deviations around the Planck 2018 constraints~\cite{PlanckCollaboration20} with each standard deviation matching \textit{Euclid} pessimistic forecast results using a combination of weak lensing and spectroscopic galaxy clustering~\cite{Euclid20}, as in Ref.~\cite{Mancarella22}. We additionally clip these ranges for $\Omega_{\textrm{b}}$, $w_0$ and $w_a$ as indicated in Table~\ref{tab:priors} to avoid numerical instabilities of CAMB.

We assume the same Gaussian noise model on the power spectra as in Ref.~\cite{Mancarella22}, which is based on the expected error for a Stage IV survey like \textit{Euclid}~\citep{Feldman94, Seo07, Zhao14}:
\begin{equation}
    \sigma(k, z) = \sqrt{\frac{4\pi^2}{k^2 \Delta k V(z)} \left( P_{\delta \delta}(k, z) + \frac{1}{\bar{n}(z)}\right)^2 + \sigma^2_{\textrm{sys}}} \ ,
    \label{eq:error}
\end{equation}
where $V(z)$ is the survey volume, $\bar{n}(z)$ represents shot noise, $\Delta k = 0.055 \ h/$Mpc is the bin width, and $\sigma_{\textrm{sys}} = 5 \ \textrm{Mpc}/h$ is a constant term added in quadrature to account for any systematic effects. The values for $V(z)$ and $\bar{n}(z)$ are reported in Table II of Ref.~\cite{Mancarella22}. The corresponding noise on the boosts $\sigma_{\mathcal{B}} (k, z)$ is obtained via linear error propagation from Eq.~(\ref{eq:boost}). We produce a total of $10^5$ boosts; we reserve $10\%$ of them to validate the model while training, and another $10\%$ to test our model after training. The remainder is used to train the machine learning model. Generating the dataset is a one-time overhead that took $\mathcal{O}(1)$ hour on a single CPU core; in the future, we will extend our framework to multiple beyond\nobreakdash-$\Lambda$CDM models by obtaining the theoretical power spectrum predictions with the efficient \texttt{ReACT} package~\cite{Bose20, Bose22}, or the corresponding emulator~\citep{SpurioMancini23}.

\begin{table}
  \centering 
  \renewcommand{\arraystretch}{1.5}
  \begin{tabular}{c c c c}
    &\textbf{Parameter} &\textbf{Min}               &  \textbf{Max}  \\
    \hline
    \hline
     \parbox[c]{-10mm}{\multirow{5}{*}{\rotatebox[origin=c]{90}{\textbf{CDM}}}} 
    &$\Omega_{\mathrm{m}}$ &  0.27 &  0.36  \\

    &$\Omega_{\mathrm{b}}$ & 0.01          &     0.08    \\

    &$h$ & 0.65          &     0.69    \\

    &$A_{\textrm{s}}$    & $2.08 \cdot 10^{-9}$                               &     $2.31 \cdot 10^{-9}$     \\    &$n_{\textrm{s}}$   & 0.93                             &     1.01     \\
    \hline
    \hline     
    \\[-2.5ex]\parbox[c]{-10mm}{\multirow{2}{*}{\rotatebox[origin=c]{90}{\textbf{Extension}}}} 
    &$w_0$ &  $-1.3$ &  $-0.7$  \\

    &$w_a$ & $-1.6$         &     $0.3$   \\
    [1.8ex]
    \hline
    \hline    
    \end{tabular}
   \caption{Minimum and maximum for the uniform ranges used to sample the cosmological parameters when generating the matter power spectra. These are centered around the Planck 2018 best-fit values~\cite{PlanckCollaboration20} with a standard deviation from a pessimistic \textit{Euclid} forecast~\cite{Euclid20}. We further limit the $\Omega_{\textrm{b}}$, $w_0$ and $w_a$ ranges to $\left[- 2.4, 1.9\right]$, $\left[-3.1, 3.1\right]$ and $\left[-5, 0.9\right]$ standard deviations, respectively, to avoid any numerical instabilities of the Boltzmann solver we use to generate the power spectra (CAMB~\cite{Lewis11}).}
  \label{tab:priors}
\end{table}

\section{Model}
\label{sec:model}
In order to obtain a compressed representation of dark energy power spectrum boosts, we employ a $\beta$-variational autoencoder ($\beta$-VAE,~\cite{Kingma13, Rezende14, Higgins17}). A schematic view of our \texttt{DE-VAE} model is reported in Fig.~\ref{fig:scheme}.

In particular, our model is made of two blocks, an encoder and a decoder, both parametrized as three layers of 1-D convolutional neural networks. After each convolutional layer, we apply the same trainable activation function as the one described in Ref.~\cite{Alsing20} and batch normalization~\cite{Ioffe15}, to make training more stable. In the encoder, the convolutional layers are followed by a dense layer with linear activation function, whose output is a set of $2d$ variables, where $d$ is the size of the so-called latent space. These variables represent the mean and the standard deviation of $d$ Gaussian distributions. We vary $d$ in our analysis from $d=1$ to $d=3$, and report the corresponding results in Sect.~\ref{sec:results}. 

We sample one point from each latent distribution, and concatenate the latent variables with the five CDM parameters (described in Sect.~\ref{sec:data} and Table~\ref{tab:priors}), to encourage the model to find a representation only of the extension to the $\Lambda$CDM model. The concatenated vector is fed through the decoder to obtain a reconstructed version $\tilde{\mathcal{D}}$ of the input data $\mathcal{D}$. The complete model is then trained by optimizing the loss function
\begin{equation}
    \mathcal{L} = \left\lvert  \frac{\mathcal{D} - \tilde{\mathcal{D}}}{\sigma_{\mathcal{B}}} \right\rvert ^2 + \beta D_{\textrm{KL}}\left[p\left(\mathbf{z}| \mathcal{D}\right)|| \mathcal{N}\left(\mathbf{0}, \mathbf{1}\right)\right] \ , 
\end{equation}
where $\left\lvert \cdot \right\rvert$ represents the $\ell^2$ norm, $D_{\textrm{KL}}\left[\cdot || \cdot\right]$ indicates the Kullback-Leibler (KL) divergence between two distributions, $\mathcal{N}\left(\mathbf{0}, \mathbf{1}\right)$ is a multivariate standard Gaussian distribution, $\beta$ is a real parameter that we tune to achieve independent latent variables, and $p\left(\mathbf{z}| \mathcal{D}\right)$ is the conditional distribution of the latent variables given the input data. We assume that $p\left(\mathbf{z}| \mathcal{D}\right)$ is the product of $d$ independent Gaussian distributions, namely
\begin{equation}
    p\left(\mathbf{z}| \mathcal{D}\right) = \prod_{i=1}^{d} p\left(z_i| \mathcal{D}\right) = \prod_{i=1}^{d} \mathcal{N}(\mu_i (\mathcal{D}), \sigma_i ( \mathcal{D}))\ ,
\end{equation}
where the means $\mu_i$ and standard deviations $\sigma_i$ are the output of the encoder given the input data $\mathcal{D}$. With this choice, the KL divergence can be computed analytically as
\begin{equation}
    D_{\textrm{KL}}\left[p\left(\mathbf{z}| \mathcal{D}\right)|| \mathcal{N}\left(\mathbf{0}, \mathbf{1}\right)\right] = \frac{1}{2} \sum_{i=1}^{d} \left(\sigma_i^2 + \mu_i^2 - 1 - 2\log{\sigma_i} \right) \ . 
\end{equation}
Our model thus has to balance the two components of the loss: while the first part aims to achieve a good reconstruction error weighted by the uncertainty on the power spectrum boost, the second piece encourages a disentangled latent space, namely a set of independent latent variables, each carrying information about the relevant parameters used to predict the boosts.

The training procedure is analogous to Refs.~\cite{SpurioMancini22, Piras23b}. We set a batch size of 256, and a starting learning rate of $10^{-3}$; we train the model until the validation loss does not improve for 50 consecutive epochs, then decrease the learning rate by a factor of 10, and repeat until the learning rate reaches $10^{-6}$. Our model is trained using the \texttt{Adam} optimizer~\cite{Kingma15}.

To quantify the dependence between the latent variables and the cosmological parameters, we estimate their mutual information (MI), an information-theoretic measure of the dependence between random variables. Given two random variables with joint probability density $p(x, y)$, their MI is defined as
\begin{equation}
    \textrm{MI}(x,y) = \int_x \int_y p(x,y) \ln \frac{p(x,y)}{p(x)p(y)} \ ,
\end{equation}
where $p(x)$ and $p(y)$ indicate the marginal distributions, and $\ln$ indicates the natural logarithm, so that MI is measured in natural units (nat). The MI between two variables is zero if and only if the two variables are independent; we refer the reader to Ref.~\citep{Vergara15} for a complete review of MI and its properties. We employ the GMM-MI library~\citep{Piras23}, which provides a robust and efficient estimator of MI and includes the uncertainty due to the finite sample size, to obtain MI estimates. We also use GMM-MI to ensure that the latent variables are disentangled: we tune $\beta$ until we reach a MI between latent variables of $\mathcal{O}(10^{-3})$ nat or less. Finally, in the case $d=1$ we further search for an explicit connection between the latent variable and the $(w_0, w_a)$ parameters using symbolic regression, which fits multiple analytic equations to the data through genetic algorithms, as implemented in the \texttt{PySR} library~\citep{Cranmer23}.

\section{Results}
\label{sec:results}
\begin{figure*}
\captionsetup[subfloat]{captionskip=-2pt}
\subfloat[$d=1$.]{%
  \includegraphics[trim={0.65cm 0.25cm 3.8cm 5cm},clip, width=\columnwidth]{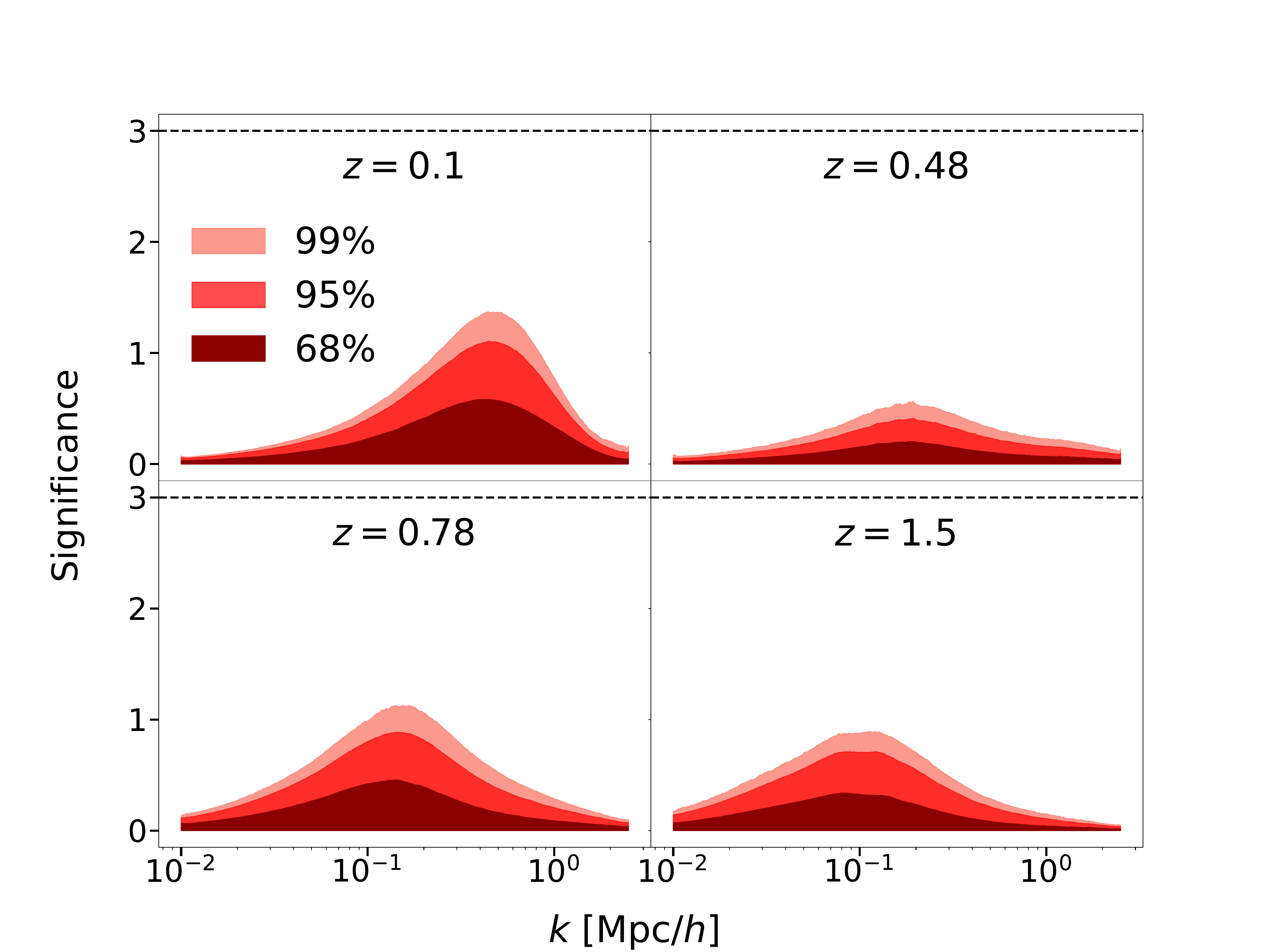}%
}
\vspace{-0.23 cm}
\captionsetup[subfloat]{captionskip=-2pt}
\subfloat[$d=2$.]{%
  \includegraphics[trim={0.65cm 0.25cm 3.8cm 5cm},clip, width=\columnwidth]{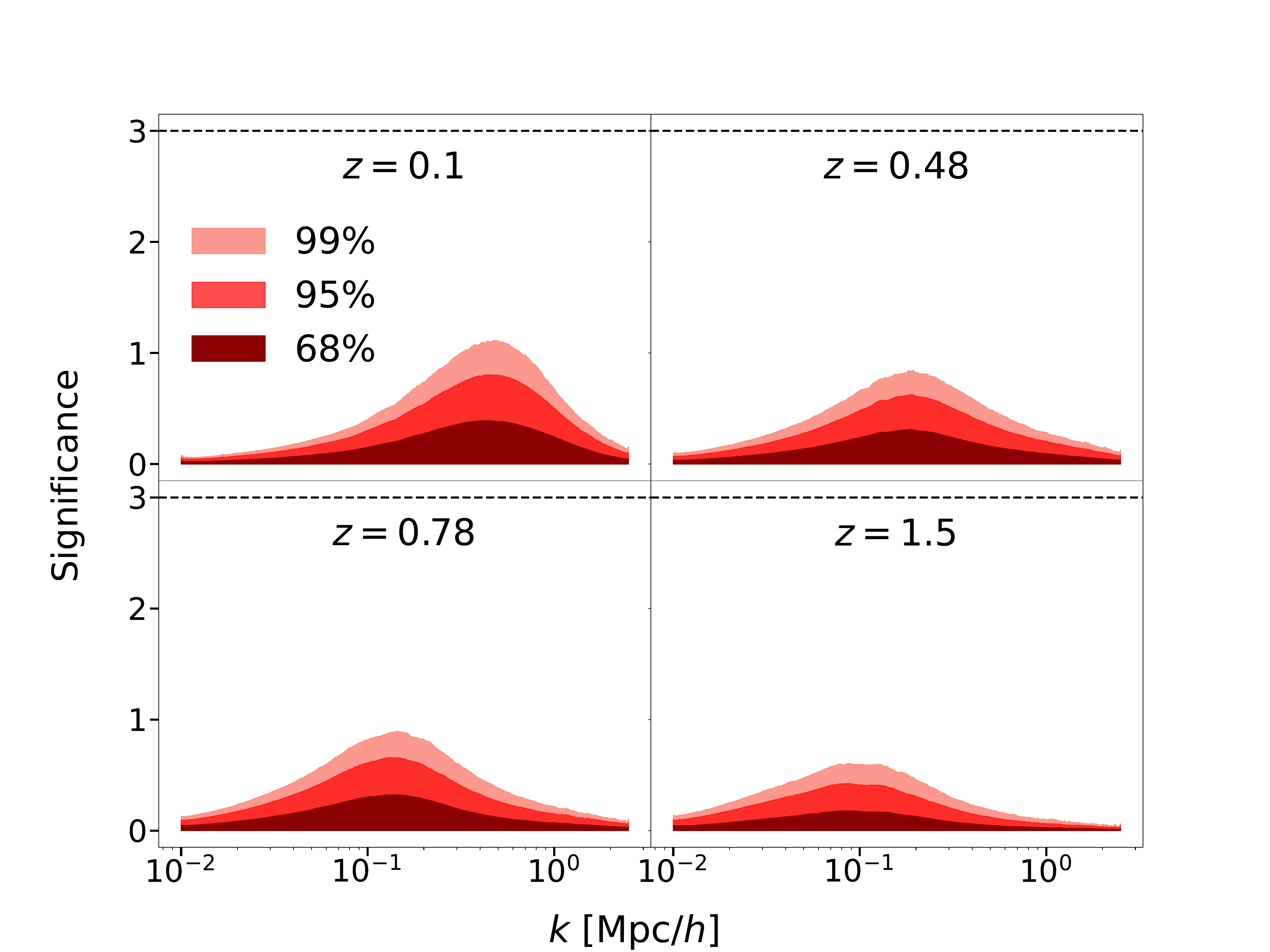}%
}
\vspace{-0.23 cm}
\captionsetup[subfloat]{captionskip=-2pt}
\subfloat[$d=3$.]{%
  \includegraphics[trim={0.65cm 0.25cm 3.8cm 5cm},clip, width=\columnwidth]{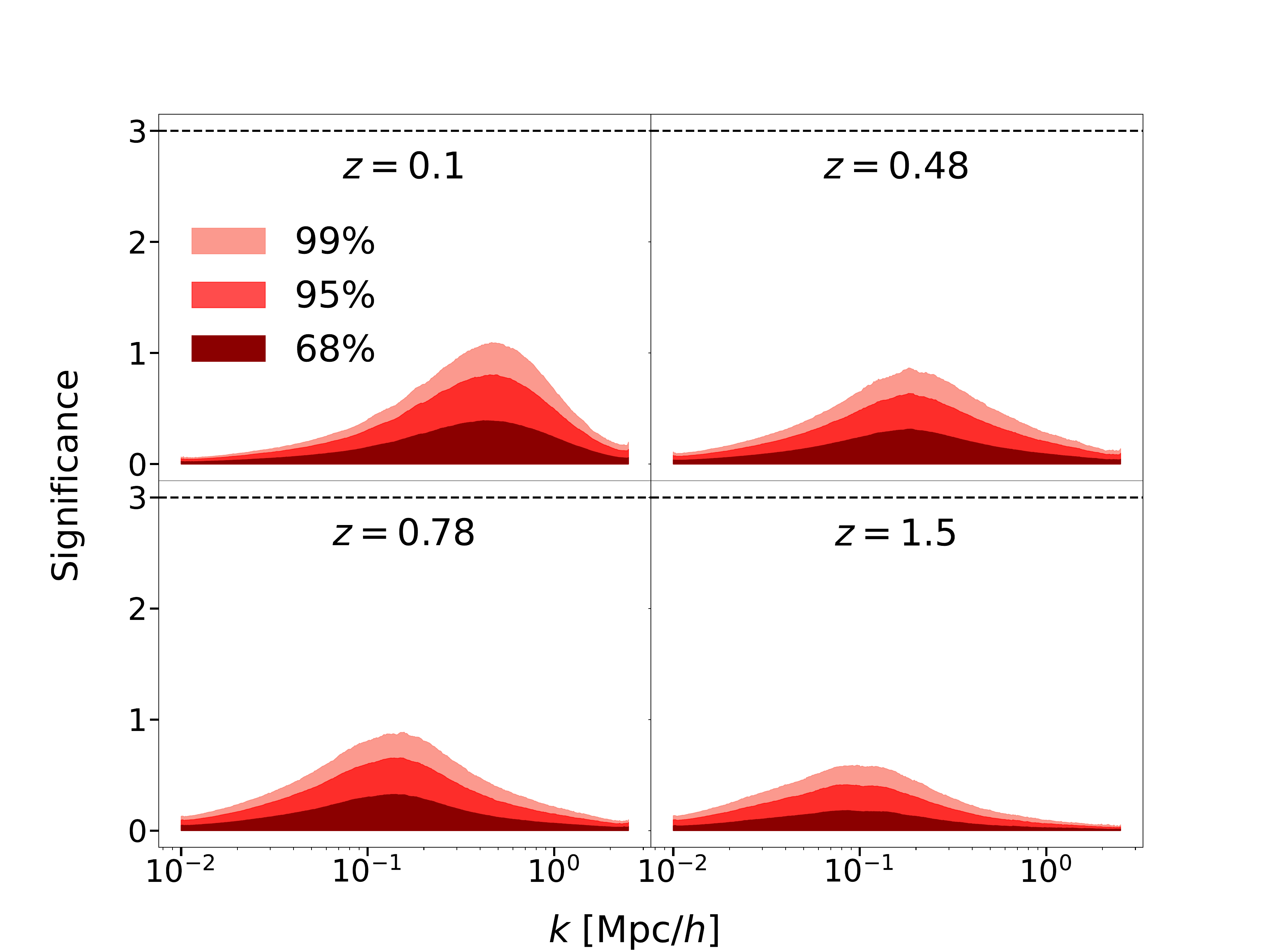}%
}
\vspace{-0.2cm}
\caption{Percentile accuracy for the \texttt{DE-VAE} model trained with different numbers of latent variables $d$. These results are obtained for wavenumbers $k \in (0.01 - 2.5) \ h/$Mpc and redshift values $z \in (0.1, 0.48, 0.78, 1.5)$. The significance is computed using Eq.~(\ref{eq:rel_diff}), which compares the test power spectra with the predictions from the model, assuming both the predicted and the ground truth power spectra have the same error as in Eq.~(\ref{eq:error}). The plots show that 99\% of the model predictions fall well within 1$\sigma$ (2$\sigma$) of the observational error for most (all) cases. The dashed horizontal lines in each panel correspond to 3$\sigma$.}
\label{fig:percentiles}
\end{figure*}

We first report the accuracy of the reconstructed test power spectra $P^{\textrm{b\nobreakdash-}\Lambda\textrm{CDM}}_{\delta\delta} (k, z)$ when varying the number of latent variables from $d=1$ to $d=3$. We compute the predicted spectrum $P^{\textrm{b\nobreakdash-}\Lambda\textrm{CDM, pred}}_{\delta\delta} (k, z)$ by multiplying the output of the decoder, considered noiseless, by $P^{\Lambda\textrm{CDM}}_{\delta\delta} (k, z)$; we then compare the predicted and true power spectra, assuming the uncertainty on the power spectra to be as in Eq.~(\ref{eq:error}). In Fig.~\ref{fig:percentiles}, we show the 68, 95 and 99 percentiles of the statistical significance,
\begin{equation}
    \rm{Significance} (k, z)=
    \frac{|P^{\textrm{b\nobreakdash-}\Lambda\textrm{CDM, pred}}_{\delta\delta} (k, z) - P^{\textrm{b\nobreakdash-}\Lambda\textrm{CDM}}_{\delta\delta} (k, z) |}{\sigma (k, z)} \ ,
    \label{eq:rel_diff}
\end{equation}
which represents the accuracy of the prediction in units of the power spectrum error across the test set, with  $\sigma(k, z)$ given by Eq.~(\ref{eq:error}). We also ensure that when $d>1$ the latent variables are disentangled, i.e.~that their MI is $\mathcal{O}(10^{-3})$ nat or less. Even in the case $d=1$, we encourage a regularized latent space by setting $\beta = 0.01$.

\begin{figure}
	\centering
	\includegraphics[width=\columnwidth]{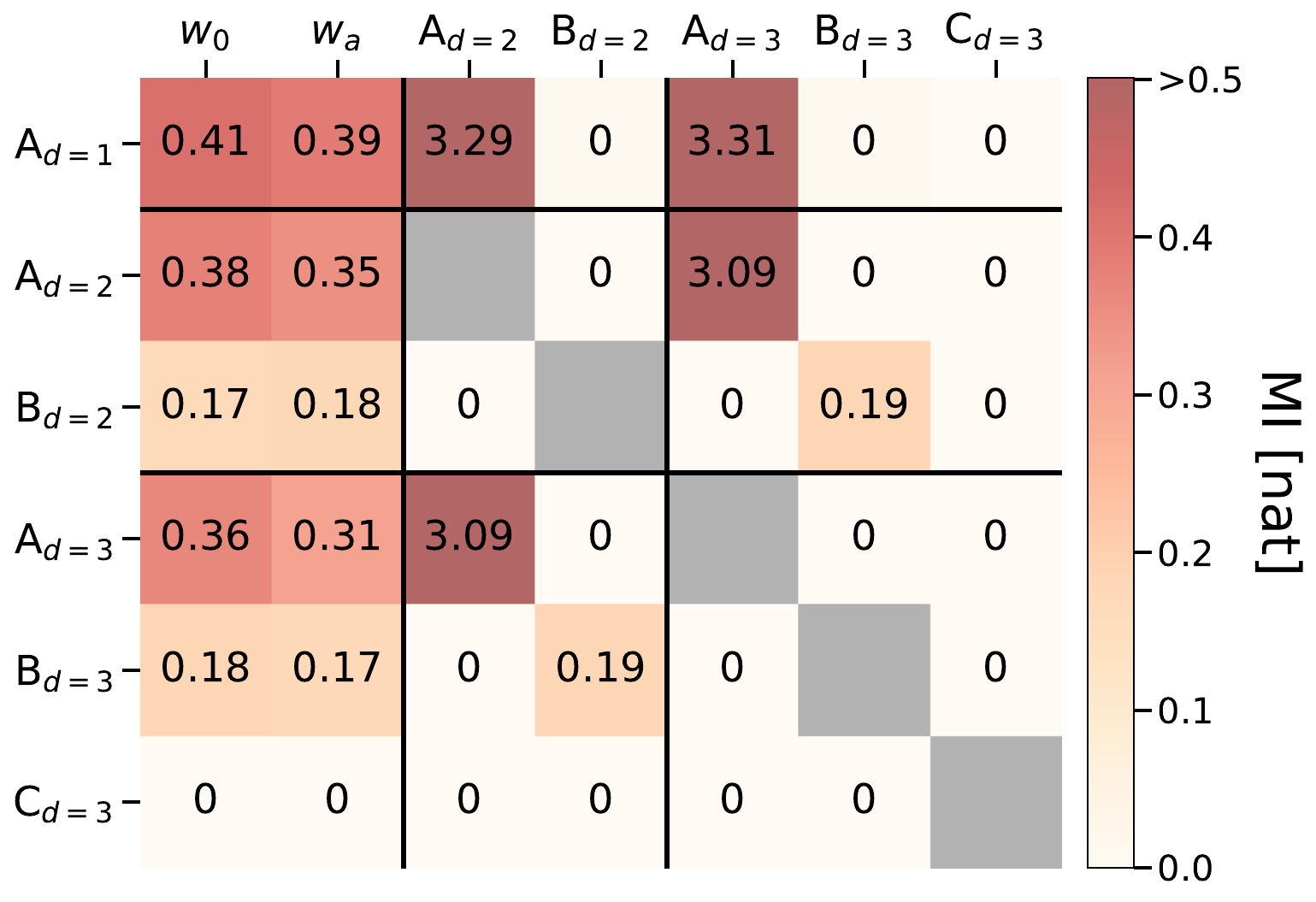}
	\caption{Mutual information (MI) between dark energy (DE) parameters $w_0$ and $w_a$ and latent variables. We consider the \texttt{DE-VAE} model trained with different numbers of latent variables $d$, and indicated with the letters A, B or C. When $d=1$, the only latent variable A$_{d=1}$ shows a high MI (i.e.~significantly higher than 0) with the DE parameters. Adding a second disentangled variable (A$_{d=2}$ and B$_{d=2}$) slightly improves the predictions of the power spectrum (as shown in Fig.~\ref{fig:percentiles}), and still shows a significant MI with the DE parameters. When $d=3$, the third latent variable (C$_{d=3}$) is essentially unused, showing no MI with either the DE parameters or any other latent variables. All values lower than 0.01 nat are indicated with a zero, while we omit the MI values in the gray squares. All MI uncertainties are negligible in this instance.}
	\label{fig:MI_values}
\end{figure} 

The trained model has a similar performance in all three cases. With one latent variable ($d=1$, top panel), the percentile plot shows that 95\% of the model predictions fall within less than 2$\sigma$ of the observational error for all wavenumbers and redshift bins, with a particularly good performance at $z=0.48$. Adding a second latent variable ($d=2$, middle panel) only marginally improves the model accuracy, which is consistent with the case of three latent variables ($d=3$, bottom panel). This suggests that adding a third independent latent variable has no impact on the model's performance: this is in line with the expectations, given that only two additional parameters are used to generate the training data.

To further interpret these results, in Fig.~\ref{fig:MI_values} we report the MI values between the latent variables, and between the latent variables and the dynamical dark energy parameters $w_0$ and $w_a$. We refer to each latent variable as A, B or C, with a subscript to indicate the dimensionality of the latent space. In the case with $d=1$, the single latent variable A$_{d=1}$ shows a high MI (i.e.~significantly higher than 0) with both $w_0$ and $w_a$, suggesting that A$_{d=1}$ carries information on a combination of the two dark energy parameters. This single variable is effectively capable of predicting dynamical dark energy power spectra, with 95\% of the predictions falling within 1$\sigma$ of the observational error, increasing to 99\% for $z>0.1$, as we showed in the upper panel of Fig.~\ref{fig:percentiles}; we further explore the links between these variables in Sect.~\ref{sec:sr}. When training a model with two disentangled latents (A$_{d=2}$ and B$_{d=2}$), these show a significant MI with both $w_0$ and $w_a$, with A$_{d=2}$ in particular capturing most of the information content. Moreover, A$_{d=2}$ has a high MI with A$_{d=1}$: this hints at the fact that, when allowing an extra latent variable, the model only learns small corrections on top of the $d=1$ model, as demonstrated by the slightly improved accuracy in the middle panel of Fig.~\ref{fig:percentiles} as well. Training a model with a third disentangled latent variable (C$_{d=3}$) shows no advantage, since C$_{d=3}$ carries no information content about the DE parameters, and is completely independent from all other latent variables. This is confirmed in the bottom panel of Fig.~\ref{fig:percentiles}, where we show that adding a third latent variable has minimal impact on the accuracy of the prediction of the model.

\subsection{Symbolic regression in latent space}
\label{sec:sr}
We further investigate the link between the dynamical DE parameters ($w_0$, $w_a$) and the single latent variable A$_{d=1}$ using symbolic regression. Without a direct link, given $w_0$ and $w_a$ one would need to create the corresponding power spectrum boost and pass it through the encoder to get the corresponding latent variable. By obtaining an explicit equation, we can both bypass the encoder (which we could achieve with another neural network, too), and reuse the parametrization in other contexts, e.g.~for studies looking at other observables beyond the power spectrum, as well as for the development of new theoretical models.

\begin{figure}
  \includegraphics[trim={1.5cm 3.5cm 2.5cm 2.9cm},clip, width=\columnwidth]{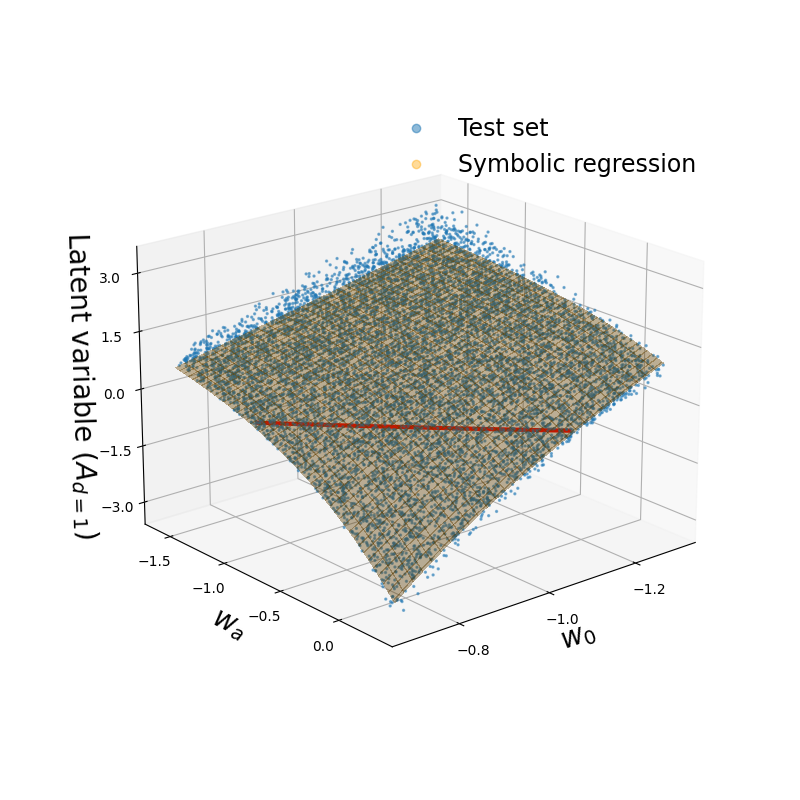}%
\caption{Scatter plot of $w_0$, $w_a$ and the $d=1$ latent variable for the test set points (blue), together with the graph of the symbolic regression equation reported in Eq.~(\ref{eq:sr}) (orange). While only accurate within the 2$\sigma$ level for 95\% of the predictions on a restricted dataset (see Fig.~\ref{fig:sr_acc}), the learned analytic expression can be used to bypass the encoder and predict the value of the latent variable given $w_0$ and $w_a$. The red line shows the intersection between the symbolic regression and $A_{d=1}=0$ planes; the latter plane describes a degeneracy with $\Lambda$CDM, which we further discuss, together with broader applications of our framework, in Sect.~\ref{sec:discuss}.}
\label{fig:3d_scatter}
\end{figure}

\begin{figure}
  \includegraphics[trim={0.75cm 0.1cm 4cm 5cm},clip, width=\columnwidth]{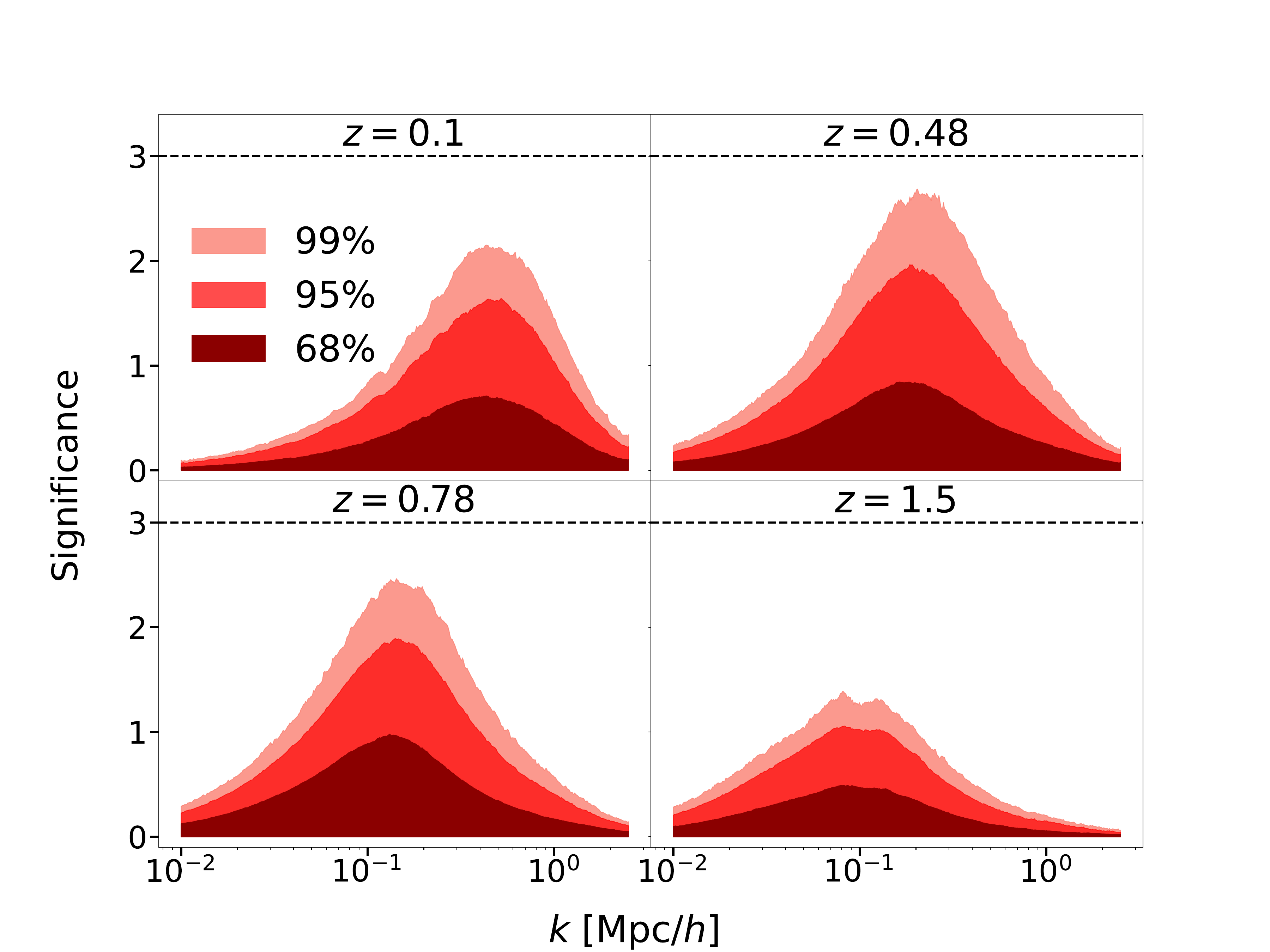}%
\caption{Same as Fig.~\ref{fig:percentiles}(a), but with the latent variable predicted using Eq.~(\ref{eq:sr}) from $w_0$ and $w_a$, and tested on 1000 boosts generated within three standard deviations of the fiducial Planck values (instead of the five standard deviations considered in the rest of this work).}
\label{fig:sr_acc}
\end{figure}

We fit \texttt{PySR} with default parameters on a subset of $10^4$ training data points; we tested that the results do not change when increasing the number of iterations, or when using more data. The best equation we obtain according to the \texttt{PySR} score (0.92 on the test set), which balances the equation's accuracy and its complexity, is
\begin{equation}
    \textrm{A}_{d=1} (w_0, w_a) = w_0^2 + \frac{e^{w_a+\cos(w_0)}}{w_0} + e^{\cos(1)} - 1  \ .
    \label{eq:sr}
\end{equation}
Note that we have manually rescaled the equation so that, for $\Lambda$CDM, $\textrm{A}_{d=1} (w_0=-1, w_a=0) = 0$. We show the 3D distribution of points for the test set, together with the graph of Eq.~(\ref{eq:sr}), in Fig.~\ref{fig:3d_scatter}. If we use this equation to predict the value of the latent variable instead of the encoder, the 99\% level of the power spectrum predictions can reach up to 5$\sigma$ deviations for the test set described in Table~\ref{tab:priors}; this is to be expected, since symbolic regression tends to sacrifice accuracy in favor of simpler expressions~\citep{Cranmer23}. However, if we restrict ourselves to spectra generated within three standard deviations of the Planck best-fit values (instead of five as the training and test data)\footnote{In this case as well, we further limit the ranges of $\Omega_{\textrm{b}}$ and $w_a$ to $\left[- 2.4, 1.9\right]$ and $\left[-3, 0.9\right]$ standard deviations, respectively, to avoid numerical instabilities, as we describe in Table~\ref{tab:priors}.}, we show that this equation actually achieves an acceptable performance, with the percentile accuracy shown in Fig.~\ref{fig:sr_acc}. We also tested that we obtain the same equation if we allow all parameters (CDM+DE) to be given as input variables to predict A$_{d=1}$, further confirming that the latent variable only depends on the DE parameters.

\section{Discussion and outlook}
\label{sec:discuss}
Using our \texttt{DE-VAE} framework, we demonstrated that a single variable is sufficient to accurately predict the theoretical matter power spectrum for a particular beyond\nobreakdash-$\Lambda$CDM model with two additional free parameters. The architecture benefits from its representation learning formulation, which allows us to explore the lower-dimensional latent space and obtain further insight on the nature of the training data. We showcased the use of mutual information as an effective tool to interpret the latent space; moreover, we employed symbolic regression (SR) to obtain an analytic equation linking the cosmological parameters with the latent variable. We envision that our framework can be employed and extended in the following ways.

\begin{itemize}
    \item Given observational constraints on the matter power spectrum, it is possible to use the \texttt{DE-VAE} decoder within a Markov chain Monte Carlo (MCMC) algorithm to sample the posterior distribution of the latent space (and CDM) parameters, and thus effectively classify the observed spectrum as either in agreement with $\Lambda$CDM or not. The value of the latent variable corresponding to $\Lambda$CDM can be obtained by passing constant unitary boosts at all redshifts and wavenumbers through the encoder. It would then be straightforward to constrain the latent parameter A$_{d=1}$ and assess the statistical significance of any deviations from $\Lambda$CDM directly in latent space. 
    \item The particular latent variable we found with \texttt{DE-VAE} and the corresponding parametrization obtained using SR are optimized for a \textit{Euclid}-like power spectrum analysis, and capture a particular degeneracy of the $w_0w_a$CDM model in predicting theoretical matter power spectra (indicated in red in Fig.~\ref{fig:3d_scatter}). We reiterate that in our framework we are not considering systematic and observational effects from which we are not necessarily independent: these would in principle require training a different VAE for each choice of nuisance parameters, and could introduce different degeneracies based on redshift, scale and other gravitational effects. On the other hand, our goal is to capture degeneracies at the theory level, both within a single model and across multiple $\Lambda$CDM extensions. The latent variable we found for $w_0w_a$CDM in this work can be interpreted in the same vein as $S_8= \sigma_8\sqrt{\Omega_{\textrm{m}}/0.3}$, i.e.~as a parameter constructed to be particularly well\nobreakdash-constrained by a certain dataset. Given that our neural architecture is easily adaptable to different input data, it will be therefore interesting to train our framework on other probes (e.g.~on CMB spectra or weak lensing maps). While this would require to produce a new dataset (possibly including additional observational effects) to train another VAE, it could yield a latent variable capturing an orthogonal degeneracy with respect to the one we found in this work, and therefore highlight which summary statistics should be targeted to break degeneracies when constraining beyond\nobreakdash-$\Lambda$CDM models. By analyzing different datasets and considering sky maps rather than 2-point functions, we aim to obtain results which are complementary with respect to the matter power spectrum, as well as retain the information that is lost when compressing the full field into summary statistics~\citep{Peel19}. In turn, this can inform the development of future cosmological missions, that could then target these particular parameters and probes to maximize their performance.
    \item When repeating the entire analysis using the \texttt{EuclidEmulator2} \citep[EE2,][]{Knabenhans21} instead of \texttt{HMCode}, we find that one latent variable is still sufficient to predict the nonlinear matter power spectra with good accuracy. While the equation found by SR when using the EE2 to generate the training data is different due to the stochasticity of the genetic algorithm, we verified that Eq.~(\ref{eq:sr}) still provides a decent fit to the mapping between $(w_0,w_a)$ and the latent variable. Moreover, we note that the degeneracy direction expressed by Eq.~(\ref{eq:sr}) is aligned with constraints obtained with the linear matter power spectrum \citep{Beauchamps21}, where it is expected that $w_0w_a$CDM models with identical linear growth factors yield indistinguishable linear power spectra. The degeneracy is then partially preserved since quasi-linear effects can be described as a function of the linear power spectrum to a good extent \citep{Mead17}. The degeneracy captured by our approach is thus robust to the choice of the nonlinear model, while being in agreement with results at the linear level. It can therefore be traced back to a theoretically well-understood phenomenological source, which proves the result to be both meaningful and interpretable. While for the particular model considered in this work this degeneracy is known, it suggests that similar (and perhaps more complex) degeneracies, which are theoretically interpretable, can be identified in the larger model space. They can then be utilized for new theoretically- and phenomenologically-informed parametrizations.
    \item By showcasing the validity of our approach with a single model for which the underlying ground truth is known, we anticipate being able to train an analogous model on a broader variety of spectra obtained assuming different beyond\nobreakdash-$\Lambda$CDM models, including Hu-Sawicki $f(R)$~\cite{Hu07} and dark sector interactions~\citep{Dvali00}, to be used as training data. The power spectrum predictions for these models can be efficiently obtained with the \texttt{ReACT} package~\cite{Bose20, Bose22, SpurioMancini23}. In this way, we expect to capture a common low-dimensional parametrization of a multitude of beyond\nobreakdash-$\Lambda$CDM scenarios; in turn, this can be used to facilitate the exploration of beyond\nobreakdash-$\Lambda$CDM models when performing MCMC analyses, and inform the theoretical development of models which try to unify these different extensions. This highlights another advantage of our framework, since performing an MCMC analysis on each of these models separately would clearly require many more iterations than running a single MCMC on the common extension found in latent space.
\end{itemize}

\section{Conclusions}
\label{sec:conclusions}
We developed \texttt{DE-VAE}, a framework to search for a compressed representation of beyond\nobreakdash-$\Lambda$CDM ($\Lambda$ cold dark matter) models using representation learning. As a first proof-of-concept study, we focused on dynamical dark energy (DE) as a straightforward extension to the $\Lambda$CDM cosmological model. In particular, we considered the archetypal two\nobreakdash-variable parametrization of the DE equation of state, and produced the corresponding matter power spectrum boosts as in Eq.~(\ref{eq:boost}). These boosts were used as training data of a $\beta$-variational autoencoder ($\beta$-VAE) model, summarized in Fig.~\ref{fig:scheme}, where we encouraged a $d$-dimensional disentangled compressed representation independent from CDM parameters.

We found that a single latent variable, in combination with five CDM parameters, can predict the power spectra of dynamical DE at different redshifts and up to $k=2.5 \ h$/Mpc within a $1\sigma$ (2$\sigma$) error for 95\% (99\%) of the data; the error, reported in Eq.~(\ref{eq:error}), is computed combining cosmic variance, shot noise and possible systematic effects for a Stage IV-like survey. This single latent variable shows a high mutual information (MI) with the parameters of the DE equation of state, confirming that the \texttt{DE-VAE} has learned a meaningful representation of the boosts. Adding a second independent latent variable only marginally improves the prediction accuracy, with both latent variables showing a significant MI with the DE parameters. Finally, we trained a model with $d=3$ and showed that the third independent latent variable has no MI with other latent variables and with the DE parameters, confirming that it has no significant impact on the prediction accuracy. In the $d=1$ case, we also demonstrated the use of symbolic regression to obtain an explicit equation linking the DE parameters and the latent variable, which can be used to facilitate the power spectrum prediction by skipping the encoder compression, and which sheds further light on the nature of the latent variable.

We highlighted a range of applications and future developments of our framework. To mention a few, these include applying \texttt{DE-VAE} to different cosmological datasets, to explore which probes should be targeted by next-generation surveys to break parameter degeneracies, as well as training the architecture on multiple extensions of $\Lambda$CDM, to shed light on their common aspects and inform the development of the underlying theories. We will explore these avenues in future work. The \texttt{DE-VAE} architecture and the data generated to train and test the model are available upon request.



\begin{acknowledgments}
We are grateful to Benjamin Bose and Benjamin Hertzsch for their feedback on this work. DP and LL were supported by a Swiss National Science Foundation (SNSF) Professorship grant (No.~202671). DP was also supported by the SNSF Sinergia grant CRSII5\_193826 ``AstroSignals: A New Window on the Universe, with the New Generation of Large Radio-Astronomy Facilities''. The computations underlying this work were performed on the Baobab cluster at the University of Geneva.
\end{acknowledgments}

\bibliography{paper}

\end{document}